\documentclass[preprint2]{aastex}
\usepackage{graphicx}
\usepackage{color}

\slugcomment{To be submitted to \emph{Astrophysical Journal}}

\begin{document}

\title{On the rotationally driven pevatron in the centre of the Milky Way.}

\author{Osmanov Z.}
\affil{School of Physics, Free University of Tbilisi, 0183-Tbilisi,
Georgia}

\author{Mahajan S.}
\affil{Institute for Fusion Studies, The University of Texas at
Austin, Austin, TX 78712, USA}

\author{Machabeli G.}
\affil{Centre for Theoretical Astrophysics, ITP, Ilia State
University, 0162 Tbilisi, Georgia}

\begin{abstract}
Based on the collective linear and nonlinear processes in a
magnetized plasma surrounding the black hole at the galactic center
(GC), an acceleration mechanism is proposed to explain the recent
detection/discovery of PeV protons. In a two stage process, the
gravitation energy is first converted  to the electrical energy in
fast growing Langmuir waves, and then the electrical energy is
transformed to the particle kinetic energy through Landau damping of
waves. It is shown that, for the  characteristics parameters  of GC
plasma, proton energy can be boosted upto $5$PeV.

\end{abstract}


%%%%%%%%%%%%%%%%%%%%%%%%%%%%%%%%%%%%%%
\section{Introduction}
%%%%%%%%%%%%%%%%%%%%%%%%%%%%%%%%%%%%%%

Recent detection of PeV protons in the Galactic Centre (GC) by the
High Energy Stereoscopic System (HESS)\citep{pevhess} has sharpened
the  focus on a major quest in high energy astrophysics- how do
elementary particles get driven to such enormous energies! It is,
perhaps, obvious that the preponderant gravitational energy in the
neighborhood may be the ultimate power source but charting the chain
of processes that channel the gravitational into particle kinetic
energy constitutes the challenge and the main objective of this
paper.

Let us begin with a short summary of the phenomenology. The analysis
of Very High energy (VHE) $\gamma$-rays, observed in the same region
as the PeV protons, shows a strong correlation between the
$\gamma$-ray distribution and location of giant gas-rich complexes
implying that the diffuse emission might have a hadronic origin
\citep{pevhess}. The HESS collaboration, for example, has observed
the diffuse VHE emission from the centre of Sagittarius (Sgr)
A$^{\star}$. The  spectrum of $\gamma$-rays (with energies up to
tens of TeV) follow a power law with a photon index $\sim 2.3$, and
as such, is a first detection of VHE photons originating in the
hadronic $pp$ interactions. It is , then, argued by the authors of
Ref. \citep{pevhess} that the parent protons, producing
$\gamma$-rays, must have energies of the order of $1$PeV . The
authors also suggest that a possible candidate for the observed PeV
protons could be Sgr A east. Although more effort is needed to interpret HESS data, the authors have come to the initial conclusion that the
acceleration rate might be of the order of $10^{37-38}$ergs
s$^{-1}$, which in turn suggests that in the past the
bolometric luminosity of Sgr A$^{\star}$ might have been by two or three orders of magnitude bigger than its currently estimated value.

Before discussing our acceleration model, it is pertinent to point
out some alternative candidates like the shock
acceleration-possibly
%might take place in the accretion flow, by means of
via the Fermi acceleration at the stand-off accretion shock
\citep{accret} or in the termination shocks of winds
\citep{lemoine} seem to be quite inadequate for catapulting
particles to such high energies \citep{pevhess,malkov}. Our model
scenario, in which a series of well-defined physical processes
conspire to accelerate protons to the observed/inferred PeV
energies, unfolds via the following two essential steps:

1) First the centrifugal force, acting differentially on the plasma
particles (on different species like electrons and protons, and
different relativistic $\gamma$ for the same species), creates
conditions in which fast growing Langmuir waves can be,
parametrically, excited. This rapid conversion of gravitational
energy into electrical energy is the first defining step of the
model \citep{langm1,langm2}.

2) Through a somewhat involved process, described in the Methods
section, these vastly amplified gravitationally driven Langmuir
waves transfer the electrical energy to particle kinetic energy
though Landau damping. The Langmuir waves are sustained by the bulk
plasma, and therefore constitute a huge reservoir of electrical
energy. The Landau damping, however, is much more selective,
operating, preferentially on the most energetic particles, imparting
them even greater energy. This is the second major step of the model
-converting the gravitationally generated electrical energy to
kinetic energy of particles.

The two-step process of energy transfer leading to enormous
acceleration of particles is most efficient when the ``impedances"
match- when the rate of growth, and of Landau 'damping of Langmuir
waves are comparable. The workability and efficiency of this overall
mechanism  has already been demonstrated in a set of papers relevant
to a variety of astrophysical settings varying from the
magnetospheres of the neutron stars (crab-like pulsars, newly born
milisecond stars-) to the vicinity of Active Galactic Nuclei (AGN)
\citep{screp1,zev,screp2}.

It has been amply shown in the above references that the
Langmuir-Landau-Centrifugal Drive (LLCD) is a phenomenally efficient
plasma mechanism that can accelerate particles to energies
$10^{18}$eV in millisecond newly born pulsars \citep{screp1,screp2},
and to $10^{21}$eV in AGN \citep{zev}.

After outlining the theoretical model, we work out, in Sec.2, the
details of the particle acceleration mechanism for typical
parameters of the galactic centre. In Sec. 3, we summarize our
results.

%%%%%%%%%%%%%%%%%%%%%%%%%%%%%%%%%%%%%%%%%%%%%%%%%%%%%%%%%%%%%%%%%%%%%%%%%%%%%%%%
\section{Theoretical Framework and Particle Acceleration}
%%%%%%%%%%%%%%%%%%%%%%%%%%%%%%%%%%%%%%%%%%%%%%%%%%%%%%%%%%%%%%%%%%%%%%%%%%%%%%%%

We will begin by giving a brief outline of the theory of
centrifugally excited Langmuir waves in a relativistic
electron-proton plasma. We will then apply this theory to to work
out an acceleration pathway to PEV energies.

\subsection{Theoretical model}

One begins with the  linearized set of equations \citep{screp1},
composed of the Euler equation
 \begin{equation}
\label{eul3} \frac{\partial p_{_{\beta}}}{\partial
t}+ik\upsilon_{_{\beta0}}p_{_{\beta}}=
\upsilon_{_{\beta0}}\Omega^2r_{_{\beta}}p_{_{\beta}}+\frac{e_{_{\beta}}}{m_{_{\beta}}}E,
\end{equation}
the continuity equation
\begin{equation}
\label{cont1} \frac{\partial n_{_{\beta}}}{\partial
t}+ik\upsilon_{_{\beta0}}n_{_{\beta}}, +
ikn_{_{\beta0}}\upsilon_{_{\beta}}=0
\end{equation}
and the Poisson equation
\begin{equation}
\label{pois1} ikE=4\pi\sum_{_{\beta}}n_{_{\beta0}}e_{_{\beta}},
\end{equation}
where ${\beta}$ is the species index, $p_{_{\beta}}$ is the first
order dimensionless momentum ($p_{_{\beta}}\rightarrow
p_{_{\beta}}/m_{_{\beta}}$), $\upsilon_{_{\beta0}}(t) \approx
c\cos\left(\Omega t + \phi_{_{\beta}}\right)$ is the zeroth order
velocity and $r_{_{\beta}}(t) \approx
\frac{c}{\Omega}\sin\left(\Omega t + \phi_{_{\beta}}\right)$ is the
radial coordinate \citep{zev}, $e_{_{\beta}}$ is the particle's
charge and $n_{_{\beta}}$ and $n_{_{\beta0}}$ are the perturbed and
unperturbed Fourier components of the number density. The first term
in the righthand side of Eq. (\ref{eul3}) is the relativistic
analogue of the centrifugal force, which, as we have already
discussed in the previous section acts on particles with different
radial coordinates and leads to the excitation of the unstable
electrostatic waves.

A little extra detail may be helpful. In an idealized version,
considering plasma to consist of two streams of protons and
electrons, one can show that the centrifugally amplified Langmuir
waves  grow at the rate \citep{zev}
\begin{equation}
 \label{grow1}
 \Gamma= \frac{\sqrt3}{2}\left (\frac{\omega_e {\omega_p}^2}{2}\right)^{\frac{1}{3}}
 {J_{\mu}(b)}^{\frac{2}{3}},
\end{equation}
where $\omega_{e,p}\equiv\sqrt{4\pi
e^2n_{e,p}/m_{e,p}\gamma_{e,p}^3}$ and $\gamma_{e,p}$ are the
relativistic plasma frequency and the Lorentz factor for the two
streams of particles, $b = \frac{2ck}{\Omega}\sin\phi_{-}$, $k$ is
the wave vector, $\phi_{e,p}$ are the phases of the corresponding
particles, $2\phi_{-} = \phi_p-\phi_e$, $J_{\mu}(x)$ is the Bessel
function of the first kind and $\mu = \omega_e/\Omega$.

\subsection{Acceleration of protons}

To make an estimate of the (relatively strong) magnetic field in
neighborhood surrounding the GC  black hole, we note that the
particle acceleration rates $\sim10^{37-38}$erg s$^{-1}$, quoted in
\citep{pevhess},  suggest that in the past the luminosity of
SgrA$^{\star}$ should have been two or three orders of magnitude
more than its currently believed value $\sim 5\times 10^{35}$erg
s$^{-1}$. Assuming equipartition of energy, one readily estimates
the local magnetic field strength to be \citep{osm7},
\begin{equation}
\label{mag} B\approx\sqrt{\frac{2L}{r^2c}}\approx
15.4\times\left(\frac{L}{5\times
10^{38}erg/s}\right)^{1/2}\times\frac{10R_{S}}{r}G,
\end{equation}
where $L$ is the bolometric luminosity of SgrA$^{\star}$,
$R_{S}=2GM/c^2$ is the Schwarzschild radius of the black hole,
$M\approx 4\times 10^6M_{\odot}$ is its mass \citep{mass} and
$M_{\odot}\approx 2\times 10^{33}$g is the solar mass. It is
straightforward to check that the Larmor radius of electrons and
protons is by many orders of magnitude less than the Schwarzschild
radius implying that the surrounding plasma is magnetized, and the
particles will, mostly, follow the field lines. Analyzing the
radio emission of Sgr A it has been revealed that the mentioned
supermassive black hole is rotating \citep{bhrot}. On the other
hand, since the rotating black hole is supposed to be spinning with
the angular velocity \citep{shapiro},
\begin{equation}
\label{rotat} \Omega\approx\frac{a c^3}{GM}\approx 2.5\times
10^{-3}\frac{a}{0.1}rad\;s^{-1},
\end{equation}
where $0\leq a\leq 1$ is a dimensionless parameter characterizing
the rate of rotation, the frozen-in condition of plasmas will,
inevitably, lead to direct centrifugal acceleration. The
acceleration becomes extremely efficient close  to the so called the
light cylinder (LC) surface defined by $R_{lc}\equiv c/\Omega$; it
is a hypothetical boundary where the linear velocity of rotation
exactly equals the speed of light.

In the present model, magnetic field lines are assumed to be
almost straight, and the centrifugal drive continues to accelerate
particles until the plasma energy density exceeds that of the
magnetic field. The acceleration process, thus, terminates when a
particle with mass $m$ achieves a Lorentz factor \citep{zev},
$$\gamma_{_{dir}}\approx\frac{1}{c}\left(\frac{e^2L}{2m}\right)^{1/3}\approx$$
\begin{equation}
\label{gdc} \approx 1.3\times 10^5\left(\frac{L}{5\times
10^{38}erg/s}\right)^{1/3}\times\left(\frac{m_e}{m}\right)^{1/3},
\end{equation}
where $m_e\approx 9.1\times 10^{-28}$g is the electron mass. Thus,
the direct centrifugally acceleration can propel protons to  a
maximum Lorentz factor of the order of $10^4$. This energy, is far
below, the proton energies detected by  HESS. But such energetic
protons form the faster of the proton component in of the  electron-
proton plasma in the accretion disc at the GC, and  comprise the
class of particles most effectively accelerated further by the
Landau damping of the Langmuir waves, a collective mode of
oscillation of the bulk plasmas. Both the single particle, and
collective mechanisms act in synch to boost the proton energies to
PeVs.

For a plasma with a wide range of electron and proton energies
(including protons with $\gamma_p\sim 10^3$) one can show that the
rate of growth $\Gamma\sim 8.8\times 10^{-4}$s$^{-1}$ is twice as
large as the rotation frequency $\Omega/2\pi$; the latter sets the
kinematic timescale. The centrifugally excited Langmuir modes,
therefore, are very efficient in extracting rotational energy.

The linear build up of the electrostatic energy is further
compounded by a nonlinear mechanism. An electrostatic wave, with a
relatively small amplitude, generates a high frequency pressure that
pushes out the particles from the perturbed area, creating low
density regions called  the caverns \citep{zakharov}. The waves
penetrate these areas, increase the high frequency pressure,
intensify the process even more and result in what has been termed
the Langmuir collapse.

Since the density perturbation is much less than the unperturbed
value, $n_0$, the corresponding change in frequency of plasmons will
be negligible as well, $\delta\omega\ll\omega$. Therefore, the
energy of plasmons (the electrostatic energy) is constant
\begin{equation}
\label{E2a} \int {d\bf r}\mid E\mid^2 = const,
\end{equation}
where $E$ is the electrostatic field.

The perturbation of density in cavities lead to the high frequency
fluctuation of pressure, $P_{hf}\approx -E^2\delta n/(24\pi
k^2\lambda_D^2n_0)$ \citep{arcimovich}, which scales as
$P_{hf}\propto E^2$, where $\delta n$ is the electron density
perturbation, $\lambda_D\equiv \sqrt{k_{B}T_e/(4\pi n_0e^2)}$ is the
Debye length scale, $k_{B}\approx 1.38\times 10^{-16}$ erg K$^{-1}$
is the Boltzmann constant, and $T_e$ is the electron temperature.

From Eq. (\ref{E2a}) it is evident that $P_{hf}\propto 1/l^q$, where
q denotes the number of relevant spatial dimensions. It is clear
that the high frequency pressure can overcome the thermal pressure,
$P_{th} = k_BT\delta n\propto 1/l^2$ only for the three dimensional
geometry \citep{zev}. We have taken into account that in cavities
the plasmons have kinetic and potential energies of the same orders
of magnitude, $k^2\lambda_D^2\sim\mid\delta n\mid/n_0$, leading to
the behaviour $\delta n\propto k^2\propto 1/l^2$.

We deduce from the physics summarized in the preceding paragraph
that inside the magnetosphere, where the plasma particles follow the
magnetic field lines implying an essentially one-dimensional ($q=1$)
kinematics, the Langmuir collapse is prohibited.

Outside the LC, however, the plasma processes are no longer defined
by rotation but predominantly by accretion. In this region the
plasma density is approximately given by \citep{zev}
\begin{equation}
\label{n} n=\frac{L}{4\eta\pi m_pc^2\upsilon R_{lc}^2}\approx
6.3\times 10^4\times \left(\frac{L}{5\times 10^{38}erg/s}\right)
cm^{-3},
\end{equation}
where $\upsilon = \sqrt{2GM/R_{lc}}$ is the velocity of the
accreting matter close to the LC zone; we have assumed that almost
$10\%$ of the rest energy of the accreting matter transforms to
radiation ($\eta = 0.1$). For the estimated number density, the
plasma frequency exceeds the cyclotron frequency: the particles
outside the LC, therefore, are not bound by the magnetic field(the
dynamics is 3D), and  a collapse might occur.

By combining the relations $\mid E\mid^2\sim 1/l^3$ and $\delta
n\sim 1/l^2$, one can show that the time behavior of the induced
electrostatic field, and the length scale of the cavern are given by
\citep{zakharov}
\begin{equation}
\label{E2} \mid E\mid\approx \mid E_0\mid\frac{t_0}{t_0-t}
\end{equation}
\begin{equation}
\label{l} l\approx l_0\left(\frac{t_0}{t_0-t}\right)^{-2/3},
\end{equation}
where $t_0$ is the collapse time scale, $E_0\approx 4\pi ne\Delta r$
is the initial electrostatic field and $\Delta r\approx
R_{lc}/(2\gamma)$ is a length scale close to LC where the
acceleration occurs \citep{zev}. We find from Eq. (\ref{E2}) that
the Langmuir collapse boosts up the initial electric field by the
factor $\left(\Delta r/l_c\right)^{3/2}$, where $l_c\approx
2\pi\lambda_D$ is the dissipation length scale;  the collapse is,
finally, terminated by means of the Landau damping. Through Landau
damping, The highly amplified electrostatic energy
($\epsilon_p\approx E^2/(8\pi n)$),
$$\epsilon_p\left(eV\right)\approx 5.3\times
10^{15}\times\frac{\eta}{0.1}\times\left(\frac{a}{0.1}\right)
^{1/2}\times$$
\begin{equation}
\label{energy1}
\times\left(\frac{T_e}{10^5}\right)^{-3/2}\times\left(\frac{10^3}{\gamma_p}\right)^5\times
\left(\frac{L}{5\times 10^{38}erg/s}\right)^{5/2},
\end{equation}
is, finally, deposited on protons through landau damping. One has to note that for optimal transfer of energy to protons, the rate of generation of Langmuir waves (measured by Γ, the instability growth rate) and the Landau damping rate, $\Gamma_{_{LD}}\approx\omega/\gamma_e^{3/2}$, should be comparable \citep{screp2}, where $\omega=\sqrt{4\pi e^2 n_p/m_p}$. One can show straightforwardly that for not violating the aforementioned condition, the minimum value for $\gamma_p$ is approximately $500$, leading to the highest achievable energies of the order of $170$PeV.

It is evident from Eq. (\ref{energy1}) that if one considers protons
with the initial Lorentz factors of the order of $10^3$, they can be
efficiently accelerated by the LLCD mechanism up to energies
recently detected by the HESS telescope.

A comment on a possible limit on maximum energies accessible to
ultra-high energy protons, imposed through interactions with soft
photons (Inverse Compton - IC), is in order. Since the associated
cooling time, $t_{_{KN}}=\epsilon_p/P_{_{KN}}$, where
\begin{equation}
\label{PKN} P_{KN}\approx \pi r_p^2m^2c^5n_{ph}(\epsilon_{ph})\mid
ln\left(4e\epsilon/m^2c^4\right)- 11/6 \mid,
\end{equation}
is the power emitted by proton per second in the Klein-Nishina
regime \citep{blum}, $r_p=e^2/m_pc^2$, $n_{ph}\approx
L/(R_{lc}^2c\epsilon_{ph})$ and $\epsilon_{ph}\sim 1$GeV, is of the
order of $10^{22}$sec, IC mechanism is not efficient in cooling
ultra-high energy particles.

Unlike IC (operative but not efficient), the curvature radiation
does not exist at all. It has already been shown that the curvature
driven current leads to generation of the toroidal magnetic field
twisting the field lines so that outside the LC the particles
followstraight trajectories. Therefore, outside the magnetosphere,
where the collapse actually takes place, it is not influenced by the
curvature energy losses.

%%%%%%%%%%%%%%%%%%%%%%%%%%%%%%%%%%%%%%%%%%%%%%%%%%%%%%%%%%%%%%%%%%%%%%%%%%%%%%%%
\section{Summary}
%%%%%%%%%%%%%%%%%%%%%%%%%%%%%%%%%%%%%%%%%%%%%%%%%%%%%%%%%%%%%%%%%%%%%%%%%%%%%%%%%%%%%
We will now describe the results as well as some detailed
description of the physics of acceleration when the LLCD mechanism
is applied to the parameters of an electron-proton plasma medium,
surrounding the black hole in the GC. Of course, the basic
motivation is  to explain the origin of PeV protons, detected,
recently, by the HESS collaboration.

Although, the relatively strong magnetic fields could provide a
direct centrifugal acceleration to particles (terminated,
eventually, by inverse Compton scattering) , the most efficient mode
of drawing energy  from the gravitational field comes, somewhat
naturally, through the exploitation of  the collective phenomena in
a plasma that allow the building and sustenance of enormous electric
fields (and large density fluctuations) in Langmuir or plasma waves.

% but the corresponding
%energies are strongly limited mainly by the breakdown of the bead on
%the wire approximation. On the second stage the relativistic
%centrifugal force leads to charge separation and parametrically
%generates the unstable Langmuir waves.

Let us now capture the essence of the working of LLCD in the GC
plasma:

1) The free-energy  available in the differential response of
different plasma particles to the gravitational field,
parametrically, drives a fastly growing linear instability in
Langmuir waves

2) In the AGN environment, the growth of wave energy is further
enhanced, nonlinearly, by what is known as  Langmuir collapse. The
physics of Langmuir collapse is such that starting from moderate
amplitudes, immense concentration of field energy (accompanied by
the creation of density cavities) results via what could be called
an explosive nonlinear instability.  The nonlinearly generated high
frequency component of pressure, further, pushes out the particles
from the cavities; the positive feedback, in turn, amplifies the
energy of the electrostatic waves.

3)This immense concentration of electrical energy (and density) is
terminated through Landau damping- the resonant feeding of particle
kinetic energy at the cost of the field energy. The nonlinear
Langmuir collapse, by decreasing the length scales of the caverns,
makes the  Landau damping process extremely efficient. At the end of
the day, the gravitational energy, through the linear and nonlinear
build up of Plasma waves and their Landau dissipation, is
efficiently transferred to protons; we show that for the following set of parameters: $T_e=10^5$K, $a = \eta = 0.1$, $\gamma_p = 10^3$, $L = 5\times 10^{38}$erg/s, the particles might reach energies of the order of $5$PeV. The minimum value of $\gamma_p=500$ when the the necessary condition, $\Gamma\sim \Gamma_{_{LD}}$ still holds, leads to the maximum proton energy of the order of $170$PeV.

%%%%%%%%%%%%%%%%%%%%%%%%%%%%%%%%%%%%%%%%%%%%%%%%%%%%%%%%%%%%%%%%%%%%%%%%%%%%%%%%
\section*{Acknowledgments}
%%%%%%%%%%%%%%%%%%%%%%%%%%%%%%%%%%%%%%%%%%%%%%%%%%%%%%%%%%%%%%%%%%%%%%%%%%%%%%%%%%%%%

The research of ZO was supported by Shota Rustaveli
National Science Foundation DI-2016-14; The work of SM
was, in part, supported by USDOE Contract No.DE-- FG 03-96ER-54366 and the research of GM was partially supported by Shota Rustaveli
National Science Foundation FR/516/6-300/14.

\end{document}